\documentclass[12pt]{article}

\usepackage{graphics}
 
\def\re{\mathop{\rm Re}\nolimits}
 
\begin{document}
 
\title{Characteristic Angles in the Wetting \\ of an Angular Region: Surface Shape}
 
\author{Yuri O.\ Popov\thanks{Corresponding author.  E-mail: {\tt yopopov@midway.uchicago.edu}} \and Thomas A.\ Witten}
\date{{\em Department of Physics, University of Chicago,} \\
{\em 5640 S.\ Ellis Ave., Chicago, IL 60637}}
\maketitle
 
\begin{abstract}

The shape of a liquid surface bounded by an acute or obtuse planar angular sector is considered by using classical analysis methods.  For acute angular sectors the two principal curvatures are of the order of the (fixed) mean curvature.  But for obtuse sectors, the principal curvatures both diverge as the vertex is approached.  The power-law divergence becomes stronger with increasing opening angle.  Possible implications of this contrasting behavior are suggested.

\end{abstract}

\begin{center} {\bf PACS}: 68.03.Cd --- Surface tension and related phenomena \end{center}


\section{Introduction}

The shape of the surface of a liquid or soap film constrained at its boundaries is a classic subject of mathematical physics~\cite{finn, landau, myshkis}.  These studies demonstrate the power of producing very smooth surfaces of controlled curvature by choosing the shape of the boundary.  The chief emphasis of these prior studies is to determine the global shape of the surface bounded by a given smooth surface.    Here we emphasize the complementary question of the local surface shape in response to a singular boundary shape: namely, a line with a sharp bend enclosing a droplet spanning a plane sector of angle $\alpha$.  The role of singularities in governing the shape and the motion of fluids has aroused great current interest.  Such singularities occur when a fluid droplet breaks apart~\cite{cohen, eggers1, brenner, shikhmurzaev2}, when it merges another droplet~\cite{eggers2, eggers3, shikhmurzaev2}, when it moves across a surface~\cite{degennes, teletzke, decker, shikhmurzaev1, mahadevan, robbins, dussan2, dussan1}, or when it moves through another fluid~\cite{belmonte, chhabra1, chhabra2, brenner}.

Surprisingly, a qualitative change in the surface shape occurs as the opening angle of the boundary $\alpha$ increases past a right angle, as we show below.  The curvatures for acute angles remain finite for the region near the vertex.  But for obtuse angles, the curvatures diverge as the vertex is approached, with a power law that varies continuously with the angle.  Similar characteristic angles (not necessarily equal to 90 degrees) are encountered in the problem of capillary rise in a vertical wedge-shaped container.  Here the meniscus height is bounded if the opening angle of the wedge is larger than some critical value and diverges as $r \to 0$ if the opening angle is less than that critical value~\cite{finn, pomeau}.  The contrasting behavior of acute and obtuse angles has also been noted for other phenomena involving Laplacian fields.  In hydrodynamic flow, the velocity field near a wedge changes qualitatively as the angle increases through a right angle~\cite{moffat}.  In diffusion (and analogous random-walk polymers) emanating into a wedge-shaped region there is a similar qualitative change of behavior~\cite{redner, considine, carslaw}.
 
Our motivation for focusing on droplets over an angular sector arises from observations of irregular droplets seen in everyday life.  These often have sector-shaped regions arising from the vagaries of deposition and substrate shape.  We have noticed that evaporation in these regions leads to distinctive drying patterns of solids dissolved in the liquid.  To understand the nature of these drying patterns requires knowledge of the surface shape.  For the circular drops the problem of the surface shape assumes very simple solution (spherical cap), allowing one to proceed with the issue of evaporation profiles up to the level of successful comparison of the theoretical results with the experimental data.  These so-called ``coffee-drop deposits'' have aroused recent interest~\cite{deegan1, deegan2, deegan3}.

\begin{figure}
\begin{center}
\includegraphics{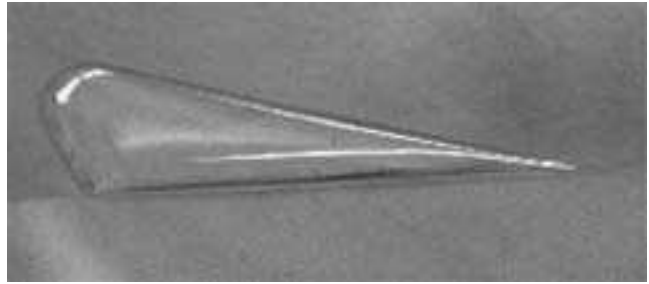}

(a)

\includegraphics{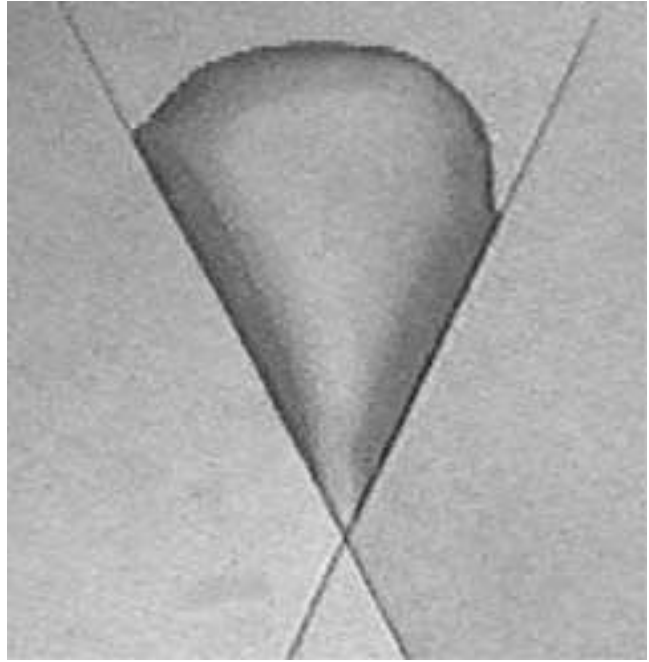}

(b)

\caption{(a) A water droplet with a sector-shaped boundary on the plane substrate (side view).  (b) The same droplet pictured from another point (top view; the experimental setup is sketched in Fig.~\ref{fig7}).  Black lines are the grooves on the substrate necessary to ``pin'' the contact line.  (Photos by Itai Cohen.)}
\label{fig1}
\end{center}
\end{figure}

\begin{figure}
\begin{center}
\includegraphics{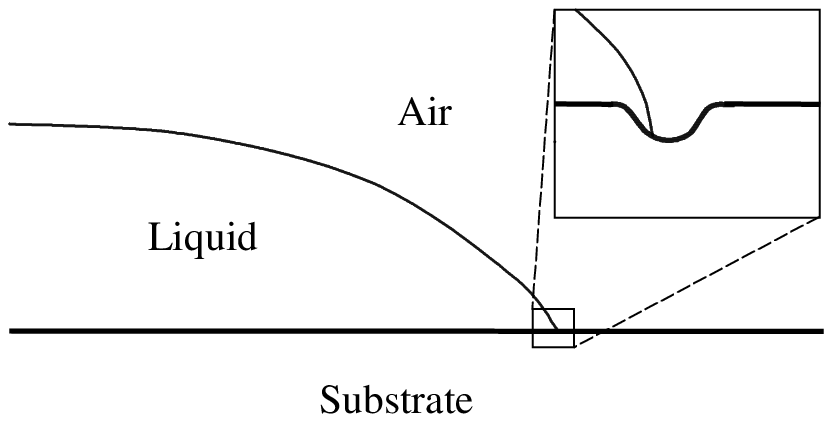}

\caption{Illustration of the possibility of a wide range of contact angles in the presence of a groove or another inhomogeneity.}
\label{fig2}
\end{center}
\end{figure}

Specifically, we consider a droplet on the horizontal surface bounded by an angle $\alpha$ in the plane of the substrate (Fig.~\ref{fig1}).  We assume that the droplet is sufficiently small so that the surface tension is dominant, and the gravitational effects can be safely neglected (significance of gravity increases with the size of the drop).  At the same time, we do {\em not\/} assume that the contact angle between the ``liquid--gas'' surface and the plane is constant along the boundary line on the substrate.  To achieve an angular boundary, the substrate must have scratches, grooves or other inhomogeneities (sufficiently small comparing to the dimensions of the droplet), which {\em pin\/} the contact line.  A strongly pinned contact line can sustain a wide range of contact angles; the angle is not fixed by the interfacial tensions as it is on a uniform surface (Fig.~\ref{fig2}).

In the following section we first give a simple account of the shape that assumes that the liquid surface is nearly horizontal, and then we make a more systematic asymptotic analysis of the region where the distance $r$ from the vertex is much smaller than the fixed inverse mean curvature $R$ of the droplet, {\em not\/} making any {\em a priori\/} assumptions about the horizontalness of the surface.  In the discussion section that follows, we calculate the curvatures of the obtained surface shape and describe some possible implications.  In particular we discuss how the refraction of the light in the drop shows contrasting properties in acute versus obtuse angular regions. 

\section{Calculation}

\subparagraph{Boundary problem.} Our purpose is to calculate the shape of the surface of the drop $z(r,\phi)$.  Use of the cylindrical coordinates looks most natural in this problem so that the angle occupied by the liquid on the substrate is $0<r<\infty$ and $-\alpha/2<\phi<\alpha/2$ and hence the boundary conditions are 
\begin{equation} z(0, \phi) = z(r, -\alpha/2) = z(r,\alpha/2) = 0 \label{boundary} \end{equation}
We start from the Laplace equation for the surface tension:  
\begin{equation} 2 H = - \frac{\Delta p}\sigma \label{laplace} \end{equation}
where $H$ is the mean curvature of the surface $H=(c_1+c_2)/2$ and $c_1$ and $c_2$ are the two principal curvatures.  Here $\sigma$ is the surface tension and $\Delta p$ is the pressure difference between liquid and gas ($\Delta p > 0$).  Since we neglect the effects of gravity, the pressure within the droplet is constant, and so is $\Delta p$.  Thus ${\Delta p}/\sigma$ is just a constant parameter of the dimensions of inverse length.  Since this is the only dimensional scale in the problem, introduction of the dimensionless variables $r / R \to r$ and $z / R \to z$ (where $R\equiv\sigma/\Delta p$) makes the mathematical formulation parameterless:
\begin{equation} 2 H = - 1 \label{dimensionless} \end{equation}
Thereby we agree to measure all quantities of the dimension of length in units of $R$.  Having found $z(r, \phi)$, one can restore the desired surface shape in ordinary units by simple substitution $z(r, \phi) \to R z\left(r/R, \phi\right)$.    

Given a surface $z(r,\phi)$, its mean curvature can be computed in terms of the coefficients of the first and the second fundamental quadratic forms of the surface:
\begin{equation} H = \frac 12 \frac{EN-2FM+GL}{EG-F^2} \label{meancurvature} \end{equation}
Here $E$, $F$, $G$ are the coefficients of the first fundamental quadratic form of the surface and $L$, $M$, $N$ are those of the second fundamental quadratic form (see \cite{korn} or \cite{finn} for a concise review of these results of the differential geometry).  For the surface $x(r,\phi) = r \cos\phi$, $y(r,\phi) = r \sin\phi$ and arbitrary $z(r,\phi)$ these coefficients are:
\begin{equation} E = 1 + z_r^2 \qquad\qquad F = z_r z_\phi \qquad\qquad G = r^2 + z_\phi^2 \label{efg} \end{equation}
and
\begin{equation} L = \frac{z_{rr}}{\sqrt{1 + z_r^2 + (z_\phi/r)^2}} \qquad M = \frac{z_{r\phi} - z_\phi/r}{\sqrt{1 + z_r^2 + (z_\phi/r)^2}} \qquad N = \frac{z_{\phi\phi} + r z_r}{\sqrt{1 + z_r^2 + (z_\phi/r)^2}} \label{lmn} \end{equation}
Combining equations~(\ref{dimensionless})-(\ref{lmn}) gives a second-order non-linear differential equation for the function $z(r, \phi)$: 
$$ \left[z_{\phi\phi} + r z_r + r^2 z_{rr} + \left\{ z_{\phi\phi} z_r^2 + r z_r^3 + z_{rr} z_\phi^2 - 2 z_{r\phi} z_r z_\phi + 2 z_r z_\phi^2 /r \right\}\right]\, + $$ 
\begin{equation} + \,r^2 \left[1 + \left\{ z_r^2 + (z_\phi/r)^2 \right\}\right]^{3/2} = 0 \label{exact} \end{equation}

Note that this equation could also have been obtained if we had tackled the problem by minimizing the surface area $A = \int\!\!\int \sqrt{1 + z_r^2 + (z_\phi/r)^2}\,r dr d\phi$ (and hence the surface energy $\sigma A$) while keeping the volume of the liquid beneath the surface $V = \int\!\!\int z\,r dr d\phi$ fixed\footnote{The integrations are over the angular region occupied by the drop.}.  This is equivalent to the minimization of the functional $A - \lambda V$ with respect to arbitrary variations of $z(r, \phi)$ that leave the boundary fixed, with $\lambda$ being a Lagrange multiplier.  The proper choice of this parameter is $\lambda = \Delta p/\sigma$ in ordinary units or $\lambda = 1$ in dimensionless ones, which arises from the expression for the total energy $E = \sigma A - V \Delta p$.  The Euler-Lagrange equation for the functional $A-\lambda V$ reads exactly as eq.~(\ref{exact}).

Thus, the boundary problem for $z(r,\phi)$ consists of the equation~(\ref{exact}) and boundary conditions~(\ref{boundary}).  Note that we do not specify the boundary conditions at the opposite side of the drop (the furthest from the vertex), and this will lead to a set of undetermined coefficients in the solution to our problem.  However, our purpose is to infer the {\em universal\/} features of the solution near the vertex, determined solely by the opening angle of the sector of interest and independent of the shape of the boundary outside of that sector.  As we show below, knowledge of this subset of boundary conditions imposes sufficiently strict limitations on possible solutions, so that many important properties of the surface shape can be determined on the basis of only these local conditions.  Had we specified all the boundary conditions, we would have obtained the exact parameter-free solution, dependent on the global shape of the boundary.

\subparagraph{Horizontal solution.} There is no generic method for solving second-order non-linear differential equations of the kind of eq.~(\ref{exact}), so we seek an approximate solution.  First of all, we notice that if all partial derivatives of $z$ are small ($|z_r| \ll 1$ and $|z_\phi /r| \ll 1$), i.e.\ if the surface is nearly horizontal, then the curly brackets in each pair of the square brackets can be neglected with respect to the rest of the terms.  This horizontal approximation is not entirely obvious, and it will be justified in the next subsection.  Thus, omitting the curly brackets in eq.~(\ref{exact}), an easy-to-solve Poisson equation is recovered:  
\begin{equation} \nabla^2 z = - 1 \label{poisson} \end{equation}
The general solution to the boundary problem~(\ref{poisson}), (\ref{boundary}) can be written as a sum of three terms:
\begin{equation} z = - \frac{r^2}4 + z_{PN} + z_{GH} \end{equation}
where $(- r^2 /4)$ is a solution to the non-homogeneous equation:
\begin{equation} \nabla^2 \left(- \frac{r^2}4\right) = - 1 \end{equation}
$z_{PN}$ is a particular solution to the homogeneous equation with non-homogeneous boundary conditions:
\begin{equation} \nabla^2 z_{PN} = 0 \qquad\mbox{and}\qquad z_{PN}(r, -\alpha/2) = z_{PN}(r,\alpha/2) = \frac{r^2}4 \label{zpn} \end{equation}
and $z_{GH}$ is the general solution to the fully homogeneous boundary problem:
\begin{equation} \nabla^2 z_{GH} = 0 \qquad\mbox{and}\qquad z_{GH}(r, -\alpha/2) = z_{GH}(r,\alpha/2) = 0 \label{zgh} \end{equation}

A particular solution to problem~(\ref{zpn}) is
\begin{equation} z_{PN}(r,\phi)=\left\{\begin{array}{ll} \frac{r^2}4 \frac{\cos 2\phi}{\cos\alpha} = \re\left(\frac{\xi^2}{4\cos\alpha}\right) &\qquad\qquad\mbox{if}\quad \alpha \ne \frac \pi 2 \\ \\ - \frac{r^2}\pi \ln r \cos 2\phi + \frac{r^2}\pi \phi \sin 2\phi = \re\left(-\frac{\xi^2 \ln\xi}\pi\right) &\qquad\qquad\mbox{if}\quad \alpha = \frac \pi 2 \end{array}\right. \end{equation}
where a complex variable $\xi = r e^{i\phi}$ has been introduced on the plane of the substrate.  Since this expression is a real part of an analytical function of $\xi$ (for each fixed $\alpha$), it is a harmonic function on $(r,\phi)$-plane by the Cauchy-Riemann conditions, and hence it is a solution to $\nabla^2 z_{PN} = 0$.  Boundary conditions can be verified by direct substitution $\phi = \pm \alpha /2$. 

The general solution to the homogeneous problem~(\ref{zgh}) must satisfy the symmetry of the problem (i.e.\ must be even in $\phi$) and can be found by standard methods of mathematical physics: 
\begin{equation} z_{GH} = \sum\limits_{n=0}^\infty C_n r^{(2n+1)\pi/\alpha} \cos\left[(2n+1)\frac{\pi\phi}\alpha\right] = \re\left(\sum\limits_{n=0}^\infty C_n \xi^{(2n+1)\pi/\alpha}\right) \end{equation}
The constants $C_n$ cannot be determined without imposing further conditions on the solution (for instance, obviously $C_n$ may depend on $\alpha$).  Had we specified the boundary conditions along some curve $r_0(\phi)$ that represents the rest of the boundary of the drop, all the $C_n$ would be fixed.  Since the number of coefficients $C_n$ is infinite, any reasonable boundary condition at $r_0(\phi)$ can be satisfied.  On the other hand, as it is apparent from our construction, those missing boundary conditions would not influence any other terms in the solution, which are universal and do not depend on the rest of the drop.

Thus, the general solution (even in $\phi$ and going to 0 as $r\to 0$ in cylindrical coordinates) to the boundary problem~(\ref{poisson}), (\ref{boundary}) is
\begin{equation} z(r,\phi)=\left\{\begin{array}{ll} - \frac{r^2}4 + \frac{r^2}4 \frac{\cos 2\phi}{\cos\alpha} + \sum\limits_{n=0}^\infty C_n r^{(2n+1)\pi/\alpha} \cos\left[(2n+1)\frac{\pi\phi}\alpha\right] &\qquad\mbox{if}\quad \alpha \ne \frac \pi 2 \\ \\ - \frac{r^2}4 + \frac{r^2}\pi \left(\phi \sin 2\phi - \ln r \cdot \cos 2\phi\right) + \sum\limits_{n=0}^\infty C_n r^{(2n+1)2} \cos\left[(2n+1)2\phi\right] &\qquad\mbox{if}\quad \alpha = \frac \pi 2 \end{array}\right. \label{flat} \end{equation}
Note the dominant terms in the limit $r \ll 1$ for different values of $\alpha$:  for acute angles ($\alpha<\pi/2$) the $r^2$-term dominates, for obtuse angles ($\alpha>\pi/2$) the $r^{\pi/\alpha}$-term does, and for the right angle ($\alpha=\pi/2$) the $(r^2 \ln r)$-term does.  At $\alpha = \pi/2$ both the $r^2$ and the $r^{\pi/\alpha}$ terms scale with $r$ as $r^2$ (i.e.\ they ``switch'' here in the sense of power dominance), and at exactly this value a logarithmic correction to $r^2$ appears, as it typically happens for a power series solution near a crossover of two powers.  

It may seem from the structure of the expression~(\ref{flat}) that this solution is a discontinuous function of $\alpha$ at $\alpha = \pi/2$.  However, this is not true.  The key observation is that the coefficients $C_n$ can be different for different values of $\alpha$.  In particular, $C_0$ in the upper line of eq.~(\ref{flat}) is not the same as the one in the lower line.  Let us keep the notation $C_0$ for the coefficient $C_0$ in the right-angle expression (lower line) and introduce a new notation $C$ for that coefficient in the expression for angles different from $\pi/2$ (upper line).  Consider some angle $\alpha$ in the vicinity of $\pi/2$, i.e.\ let $\alpha=\pi/2+\varepsilon$, where $|\varepsilon| \ll 1$, and expand the second and the $C r^{\pi/\alpha} \cos(\pi\phi/\alpha)$ terms in the result for $\alpha \ne \pi/2$ in small parameter $\varepsilon$:  
$$\frac{r^2}4 \frac{\cos 2\phi}{\cos\alpha} + C r^{\pi/\alpha} \cos(\pi\phi/\alpha) = \frac{r^2}4 \left(-\frac{\cos 2\phi}\varepsilon\right) \left(1+O(\varepsilon^2)\right) \, + $$
\begin{equation} + \, C r^2 \left(1- \frac 4\pi \varepsilon \ln r + O(\varepsilon^2)\right)\left(\cos 2\phi + \frac 4\pi \varepsilon \phi \sin 2\phi + O(\varepsilon^2)\right) \label{expansion} \end{equation}
Now, since $C$ can depend on $\alpha$ (and hence $\varepsilon$), we set
\begin{equation} C = \frac 1{4\varepsilon} + C_0 + O(\varepsilon) \label{c} \end{equation}
Then the two diverging terms of the order of $1/\varepsilon$ cancel, and we recover (up to the leading order in $\varepsilon$) the second and the $C_0 r^{\pi/\alpha} \cos(\pi\phi/\alpha)$ terms in the result for $\alpha = \pi/2$:
\begin{equation} \frac{r^2}4 \frac{\cos 2\phi}{\cos\alpha} + C r^{\pi/\alpha} \cos(\pi\phi/\alpha) = - \frac{r^2}\pi \ln r \cos 2\phi + \frac{r^2}\pi \phi \sin 2\phi + C_0 r^2 \cos 2\phi + O(\varepsilon) \end{equation}
Since the other terms in the result~(\ref{flat}) are identical in the upper and the lower lines, we have shown thereby that our solution is indeed continuous in $\alpha$ at fixed $r$ for $\alpha = \pi/2$ (or $\varepsilon = 0$).  

Thus, solution~(\ref{flat}) behaves reasonably well in the full range of values of angle $\alpha$ from 0 to $\pi$.  It cannot be used for $\alpha > \pi$ since in that range it violates the horizontalness requirement employed in its derivation.

\subparagraph{Asymptotic analysis.}  The results above required the assumption that the drop is nearly horizontal.  This assumption has not been justified yet, and now we justify it via a more systematic treatment.  Since we are interested in the behavior of the surface {\em near\/} the vertex of the angle, we introduce a new small parameter for the problem \begin{equation} r \ll R = \frac{\sigma}{\Delta p} \qquad\qquad\mbox{(ordinary units)}  \label{condition0} \end{equation}
or
\begin{equation} r \ll 1 \qquad\qquad\mbox{(dimensionless units)}  \label{condition} \end{equation}
For small $r$ we may write $z(r, \phi)$ as a standard series expansion:
\begin{equation} z(r,\phi)=r^\nu \Phi_\nu (\phi) + r^\mu \Phi_\mu (\phi) + \cdots \label{series} \end{equation}
where $0<\nu<\mu<\ldots$.  Note that we do not restrict our attention to the horizontal case only, i.e.\ we do not require $1 < \nu$.  Values of $\nu$ between 0 and 1 leading to non-horizontal surfaces will be eliminated automatically by application of boundary conditions to the solutions of eq.~(\ref{exact}), thus justifying the horizontal assumption.  Here we find only the main asymptotic ($\nu$-term) and the first order correction ($\mu$-term), but the method allows one to proceed up to an arbitrary order.  Details of the calculation are considered in the Appendix; results are presented below.  We also treat the case of the right angle separately since we expect logarithmic corrections to the main power of $r$ and failure of the assumption~(\ref{series}). 

\subparagraph{Leading asymptotic.}  Substitution of $z(r,\phi)=r^\nu \Phi_\nu (\phi)$ into eq.~(\ref{exact}) and retention of only the dominant terms for $r \ll 1$ lead to different equations for different possible values of power $\nu$.  Solution of those equations and application of symmetry arguments and boundary conditions eliminate some\footnote{The leading-order angular solution for $\nu \le 1$ is $\Phi_\nu(\phi) = C \cos^\nu \phi$.  This cannot vanish as required for $\phi \to \pm \alpha/2$ (see Appendix for details).} values of $\nu$, leaving at the end only two possibilities ($\nu=2$ and $\nu=\pi/\alpha $).  For these two values the terms retained are a subset of those constituting eq.~(\ref{poisson}), yielding the following main order result:
\begin{equation} z(r,\phi)=\left\{\begin{array}{lll} \frac 14 r^2 \left(\frac{\cos 2\phi}{\cos\alpha}-1\right) &\qquad\qquad\mbox{if}\quad 0\le\alpha<\frac \pi 2 &\qquad(\nu=2)\\ \\ C r^{\pi/\alpha} \cos\frac{\pi\phi}{\alpha} &\qquad\qquad\mbox{if}\quad \frac \pi 2 <\alpha\le\pi &\qquad(\nu=\pi/\alpha) \end{array}\right. \label{solution0} \end{equation}
This agrees with the leading behavior of the horizontal solution~(\ref{flat}) as $r \to 0$.  Thus, our surface is indeed nearly horizontal (since $\nu > 1$), and the horizontal approach indeed produced a sensible result.  

The constant $C$ is again restricted by neither the equation nor the side boundary conditions, but it would get fixed once the boundary conditions at the furthest side are taken into account.  It is a direct equivalent to the constant $C$ in the horizontal solution~(\ref{flat}) obtained by an independent treatment (recall that we relabeled the $C_0$ in the upper line of eq.~(\ref{flat}) into $C$).

\begin{figure}
\begin{center}
\includegraphics{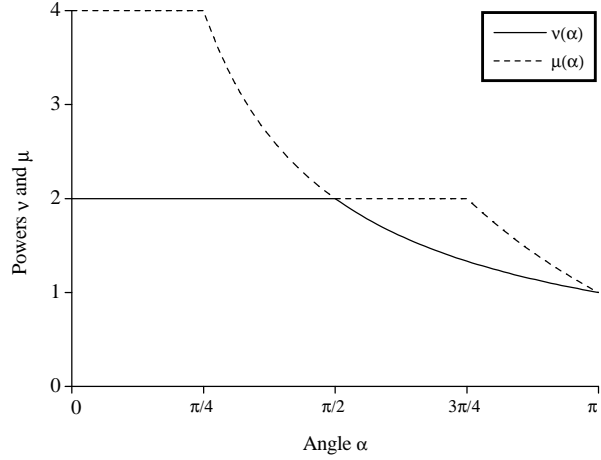}

\caption{Dependences $\nu(\alpha)$ and $\mu(\alpha)$.}
\label{fig3}
\end{center}
\end{figure}

Dependence $\nu(\alpha)$ is shown on Fig.~\ref{fig3}:  for acute angles $\nu=2$ and for obtuse ones $\nu=\pi/\alpha$.  

\subparagraph{First order correction.}  The final first-order result for the surface shape in the limit $r \ll 1$ is:
\begin{enumerate}
\item
If $0<\alpha<\pi/4$, then $\nu=2$, $\mu=4$ and 
$$ z(r,\phi)=\frac 14 r^2 \left(\frac{\cos 2\phi}{\cos\alpha}-1\right)\, + $$
\begin{equation} +\,\frac 1{192} r^4 \left(\left[4\tan^2\alpha-3\right] \frac{\cos 4\phi}{\cos 2\alpha} + \left[2\tan^2\alpha+12\right] \frac{\cos 2\phi}{\cos\alpha} - \left[6\tan^2\alpha+9\right]\right) + \cdots \label{solution1} \end{equation}
\item
If $\pi/4<\alpha<\pi/2$, then $\nu=2$, $\mu=\pi/\alpha$ and 
\begin{equation} z(r,\phi)= \frac 14 r^2 \left(\frac{\cos 2\phi}{\cos\alpha}-1\right) + C r^{\pi/\alpha} \cos\frac{\pi\phi}\alpha + \cdots \label{solution2} \end{equation}
\item
If $\pi/2<\alpha<3\pi/4$, then $\nu=\pi/\alpha$, $\mu=2$ and 
\begin{equation} z(r,\phi)= C r^{\pi/\alpha} \cos\frac{\pi\phi}\alpha + \frac 14 r^2 \left(\frac{\cos 2\phi}{\cos\alpha}-1\right) + \cdots \label{solution3} \end{equation}
\item
If $3\pi/4<\alpha<\pi$, then $\nu=\pi/\alpha$, $\mu=3\pi/\alpha-2$ and 
\begin{equation} z(r,\phi)= C r^{\pi/\alpha} \cos\frac{\pi\phi}\alpha + C^3 \frac{\pi^3}{4\alpha^2(2\pi-\alpha)} r^{(3\pi/\alpha) -2} \cos\frac{\pi\phi}\alpha + \cdots \label{solution4} \end{equation}
\end{enumerate} 
The structure of the solution~(\ref{solution1})-(\ref{solution4}) becomes clear if one plots functions $\nu(\alpha)$ and $\mu(\alpha)$ together (Fig.~\ref{fig3}).  Four powers of $r$ appear in these formulas: $r^2$, $r^4$, $r^{\pi/\alpha}$ and $r^{(3\pi/\alpha)-2}$.  For any given $\alpha$ our procedure selects the two lowest powers in this set of four.  Different powers get selected for different $\alpha$; this leads to the four cases appearing in (\ref{solution1})-(\ref{solution4}).  In a full expansion, we expect {\em all\/} four powers to be present for all angles\footnote{Note that all terms satisfy boundary conditions independently.}.  

Since a plane $z = D x = D r \cos \phi$ is the {\em exact\/} solution to the equation~(\ref{exact}) when the boundary of the surface is a straight line $\alpha=\pi$, it is rewarding to observe that both terms in the expression~(\ref{solution4}) reduce to this functional form with $D = C + \left(C^3/4\right) + \cdots$ as $\alpha\to\pi$.

Note that although the leading asymptotics in $r$ is the same in the results~(\ref{flat}) and (\ref{solution1})-(\ref{solution4}), the sub-leading terms are different.  This is due to the fact that the two results are based on different approximations:  the former assumes $|z_r| \ll 1$ and $|z_\phi /r| \ll 1$ while the latter assumes $r \ll 1$.  Since, as shown in the previous subsection, the horizontal approximation follows from the close-to-the-vertex one ($r \ll 1$), the asymptotic treatment of this subsection describes the surface shape more accurately than result~(\ref{flat}), picking up lower powers of $r$ for the first sub-leading terms.

\subparagraph{Right-angle sector: first two terms in the expansion.}  We already know that a pure power series does not work in the case $\alpha=\pi/2$ and that the leading power of $r$ should be close to 2, at most logarithmically close.  So, we introduce a new {\em ansatz\/} instead of series~(\ref{series}):
\begin{equation} z(r,\phi)=(-r^2 \ln r) \Phi_1 (\phi) + r^2 \Phi_2 (\phi) + \cdots \label{log-series} \end{equation}
Subsequently, by repeating the steps of the Appendix, we find $\Phi_1(\phi)=A\cos 2\phi$, then obtain the second term ($\Phi_2(\phi)$) and fix $A=1/\pi$ by boundary conditions.  As a result we recover the lower line of the expression~(\ref{flat}):
\begin{equation} z(r, \phi)= - \frac 1\pi r^2 \ln r \cos 2\phi + r^2 \left( \frac 1\pi \phi \sin 2\phi - \frac 14 + C_0 \cos 2\phi \right) + \cdots \label{right-angle0} \end{equation}
where $C_0$ is again a constant equivalent to the $C_0$ in~(\ref{flat}).  The relation between the constant $C_0$ in this right-angle expression and the constant $C$ in the result~(\ref{solution1})-(\ref{solution4}) for $\alpha\ne\pi/2$ is exactly the same as in eq.~(\ref{c}) due to the continuity of the full solution for all values of $\alpha$. 

\vspace{3ex}

Thus, an asymptotic expansion of the dimensionless function $z$ of the dimensionless coordinates $r$ and $\phi$ in the limit $r \ll 1$ has been found for all values of $\alpha$ in the range from 0 to $\pi$.  A natural extension of this study is to consider angular regions that have $\alpha > \pi$.  We guess that for such cases $z(r,\phi)$ grows as a power of $r$ less than unity, so that the slope of the surface diverges at the vertex.  But we have not succeeded to verify this behavior with our methods.

\begin{figure}
\begin{center}
\includegraphics{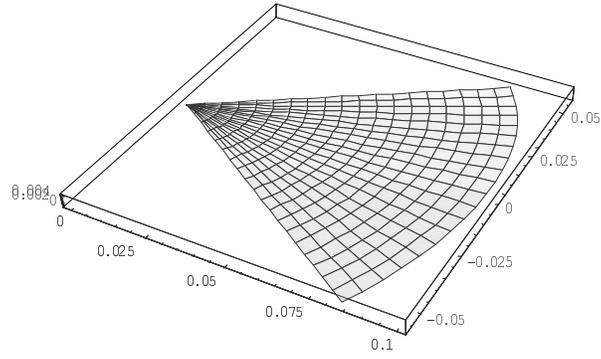}

(a)

\vspace{1ex}

\includegraphics{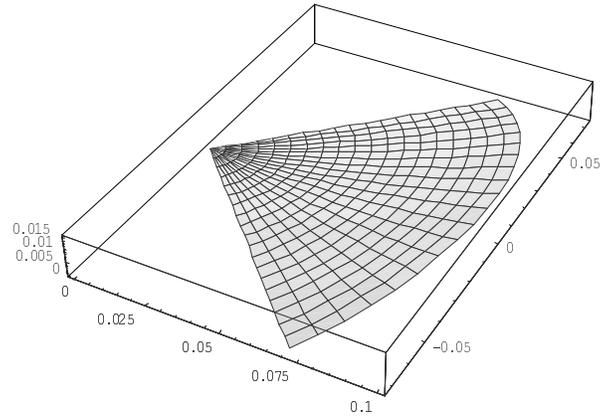}

(b)

\vspace{1ex}

\includegraphics{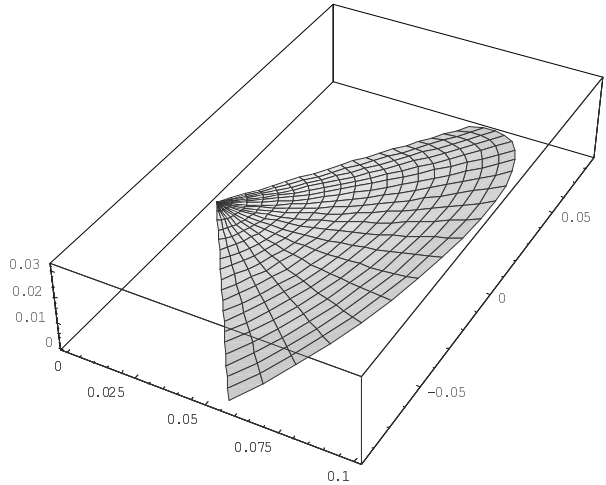}

(c)

\vspace{1ex}

\caption{Surface shape $z(r,\phi)$ (first two terms of the expansion in $r$ with $C=1/(4\alpha-2\pi)+C_0$ and $C_0=1$) for (a) $\alpha =3\pi/8$ (eq.~(\ref{solution2})), (b) $\alpha=\pi/2$ (eq.~(\ref{right-angle0})), and (c) $\alpha=5\pi/8$ (eq.~(\ref{solution3})).  The mathematical differences in the shapes are not apparent in this view.}
\label{fig4}
\end{center}
\end{figure}

\begin{figure}
\begin{center}
\includegraphics{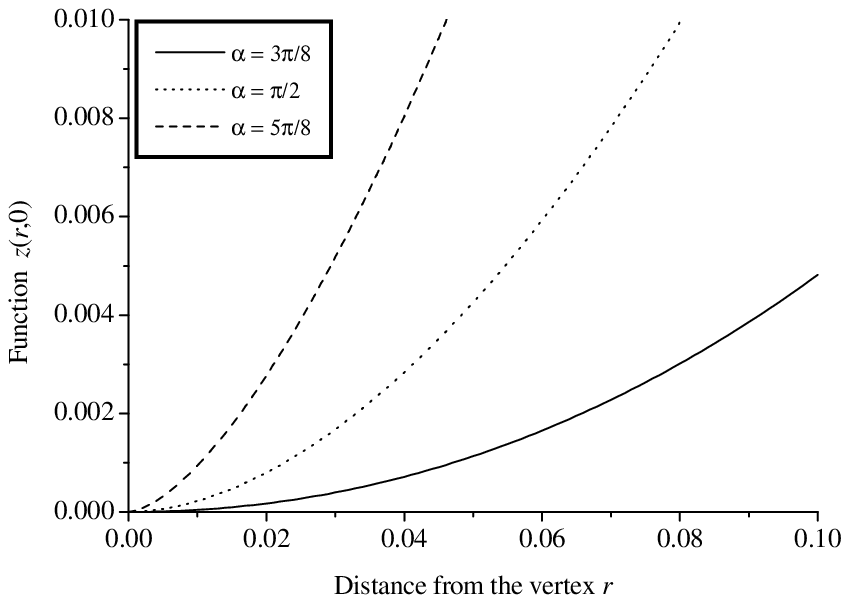}

\caption{Surface profiles at bisector ($\phi=0$) for three values of angle $\alpha$.}
\label{fig5}
\end{center}
\end{figure}

Typical behavior of the universal function $z(r, \phi)$ (eqs.~(\ref{solution1})-(\ref{solution4}) and (\ref{right-angle0})) for angles $\alpha=3\pi/8$, $\pi/2$ and $5\pi/8$ is shown on Fig.~\ref{fig4}.  To facilitate the comparison, we also plot the bisector cross-sections of the same three surface profiles in one frame (Fig.~\ref{fig5}).  Constant $C_0$ is taken to be 1 in all cases, while the value of constant $C$ is chosen according to the prescription~(\ref{c}).  In spite of the existence of drastic mathematical difference between the three regimes, this difference is not apparent from looking at the profile of the surface itself (as on Fig.~\ref{fig4}), and that is probably why it went unnoticed so far.  We will further emphasize this qualitative difference in the following section.  

\section{Discussion}

\subparagraph{Curvature.}  Let us better understand the main order result by looking at the curvature of the surfaces described above.  The principal curvatures along the bisector of the angular region are in the $\hat r$ and $\hat\phi$ directions, and for the points on the bisector the radial ($c_{rr}$) and the azimuthal ($c_{\phi\phi}$) curvatures are simply\footnote{For an arbitrary point of the surface the principal curvatures are determined by the roots $c_1$ and $c_2$ of the quadratic equation
$$\left| \begin{array}{cc}
L - c E & M - c F \\ M - c F & N - c G
\end{array} \right| = 0$$
At the points where the principal directions are along the coordinate ones (e.g.\ on bisector) coefficients $F$ and $M$ vanish and the principal curvatures are given by simple relations~(\ref{crrcphiphi}) (see~\cite{finn, korn} for details).}
\begin{equation} c_{rr} = \frac L E \qquad\qquad\mbox{and}\qquad\qquad c_{\phi\phi} = \frac N G \label{crrcphiphi} \end{equation}
with $E$, $G$, $L$ and $N$ defined in eqs.~(\ref{efg})-(\ref{lmn}).  For small $r$ (and therefore small $|z_r|$ and $|z_\phi /r|$) these expressions simplify even further to 
\begin{equation} c_{rr} = z_{rr} \qquad\qquad\mbox{and}\qquad\qquad c_{\phi\phi} = \frac{z_{\phi\phi}}{r^2} + \frac{z_r}{r} \end{equation}
so that the principal curvatures sum to $\nabla^2 z$.  Thus, for the surface~(\ref{solution0}) and (\ref{right-angle0}) (up to the leading order in $r \ll 1$ only) these principal curvatures are:
\begin{equation} \begin{array}{lll} 
c_{rr} = - \frac 1 2 + \frac 1 {2\cos\alpha} & \qquad\qquad c_{\phi\phi} = - \frac 1 2 - \frac 1 {2\cos\alpha} & \qquad\qquad\mbox{if}\quad 0 \le \alpha < \pi/2 \\ \\
c_{rr} = - \frac 2 \pi \ln r & \qquad\qquad c_{\phi\phi} = \frac 2 \pi \ln r & \qquad\qquad\mbox{if}\quad \alpha = \pi/2 \\ \\
c_{rr} = C \frac {\pi(\pi-\alpha)}{\alpha^2} r^{(\pi/\alpha) - 2} & \qquad\qquad c_{\phi\phi} = - C \frac {\pi(\pi-\alpha)}{\alpha^2} r^{(\pi/\alpha) - 2} & \qquad\qquad\mbox{if}\quad \pi/2 < \alpha \le \pi 
\end{array} \label{curvature} \end{equation} 

\begin{figure}
\begin{center}
\includegraphics{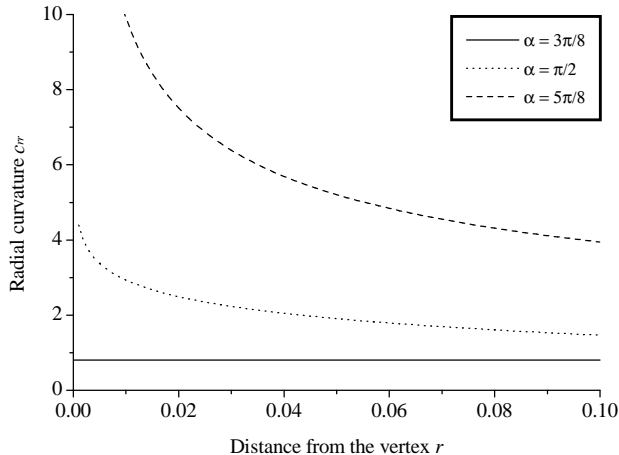}

\caption{Radial curvature on the bisector for three values of angle $\alpha$ (main asymptotic only).}
\label{fig6}
\end{center}
\end{figure} 

Typical behavior of radial curvature $c_{rr}$ on the bisector as a function of $r$ is shown on Fig.~\ref{fig6} for the same three values of $\alpha$ as on Figs.~\ref{fig4} and \ref{fig5}.  Now the dramatic difference between the two regimes of $\alpha$ becomes apparent.  For acute angles the curvatures remain finite as $r \to 0$ while for obtuse ones the curvatures diverge as a negative power of $r$ (changing from 0 to $-1$ as angle $\alpha$ passes from $\pi/2$ to $\pi$).  The limiting case of the right angle has an intermediate logarithmic divergence.  

Note that the finite values of curvature (for $\alpha < \pi/2$) sum to $-1$ in full accord with equation~(\ref{dimensionless}).  However, the divergent values (for $\alpha\ge\pi/2$) have the opposite signs and thus sum to 0.  This is a result of the neglect of the corrections to the main asymptotic in $r$.  Had we kept the corrections to the divergent curvatures, they would sum to $-1$.  Thus, for instance, the second term in the solution~(\ref{solution3}) is the kind of the correction which provides summation to $-1$ when the main (divergent) terms sum to 0.   

The origin of the arbitrary constant $C$ in the solution~(\ref{solution0}) is now seen to be related to the divergence of principal curvatures for obtuse angles.  Indeed, for acute angles the curvatures are finite and $c_{rr} + c_{\phi\phi} = -1$ while for the obtuse ones the curvatures are divergent and $c_{rr} + c_{\phi\phi} = 0$ (up to the main order in $r$).  Therefore, multiplication of the solution $z(r,\phi)$ by an arbitrary constant is not allowed in the former case while is perfectly legitimate in the latter one (since both $c_{rr}$ and $c_{\phi\phi}$ get multiplied by the same constant).  Thus, a possibility for an arbitrary multiplicative constant in the solution for obtuse angles comes from the divergence of the curvature, which in its turn reflects a different $r$-dependence of $z(r,\phi)$ for $\alpha > \pi /2$.  

On the other hand, the presence of an undetermined coefficient means that the shape is influenced by the boundary conditions at the side of the drop furthest from the vertex.  Hence, different scaling with $r$ for different angles results in different dependence on boundary conditions outside of the sector of interest:  for acute angles the dimensionless shape of the surface at the tip of the sector does not depend on these boundary conditions while for obtuse angles it does.  Of course, this argument was based on the main-order solution~(\ref{solution0}), but general dependence of the surface shape on the boundary conditions at the opposite side of the drop via constant $C$ is already apparent from the first-order solution~(\ref{solution1})-(\ref{solution4}): the larger the opening angle, the stronger dependence of $z(r,\phi)$ on these yet-unspecified boundary conditions (the number of terms containing $C$ increases as $\alpha$ increases).  This seems quite reasonable, as intuitively surface shape near the vertex must cease depending on the rest of the drop as $\alpha \to 0$, and it must be fully specified by the rest of the drop when there is no vertex at all (i.e.\ when $\alpha = \pi$). 

\begin{figure}
\begin{center}
\includegraphics{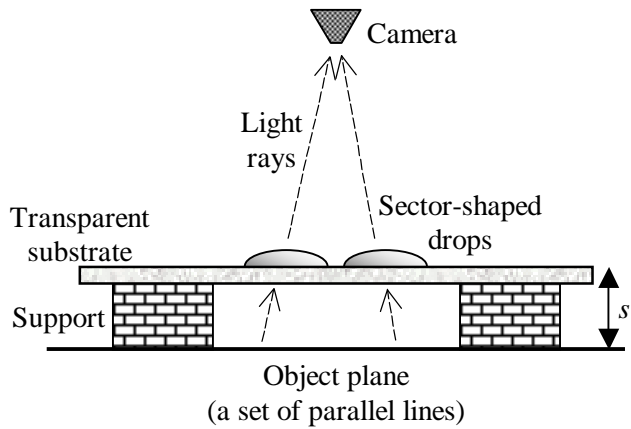}

\caption{Experimental setup for the refraction demonstration.}
\label{fig7}
\end{center}
\end{figure}

\begin{figure}
\begin{center}
\includegraphics{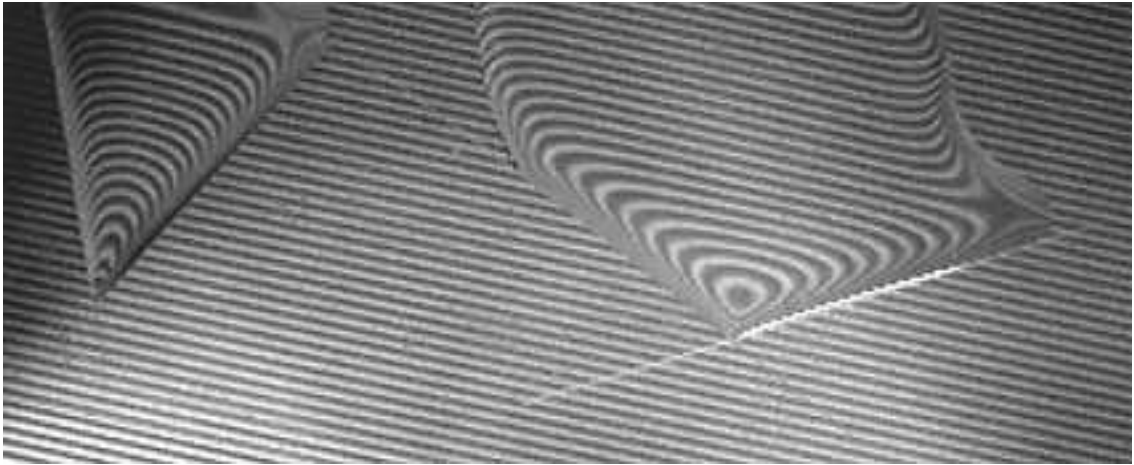}

\caption{A set of parallel lines as seen through the drops over acute (left) and obtuse (right) angular regions.  The spacing between the undistorted lines is about $1.6$~mm.  The opening angles of the acute and obtuse sectors are approximately $51^{\circ}$ and $124^{\circ}$ respectively.  The regions of interest are only a few lines (a few millimeters) around the vertex of each drop.  The grooves necessary to pin the contact line can also be seen on this image.  (Photo by Itai Cohen.)}
\label{fig8}
\end{center}
\end{figure}

\subparagraph{Experimental realizations.}  The contrast between the drops over acute and obtuse angular regions may be seen in the way they refract light.  To illustrate, we picture a pair of sector-shaped droplets on a transparent substrate at a distance $s$ above an object plane (Fig.~\ref{fig7}).  One angle is acute, and the other one is obtuse.  The object plane consists of a set of closely spaced parallel lines perpendicular to the bisectors of each angular region, so that the spacing between the lines is in the $\hat r$ direction along the bisector.  Observation of the object plane through the droplets allows one to make qualitative judgment about the behavior of the curvature near the vertices of each sector.  The result of such a simple demonstration with spacing between the parallel lines of approximately $1.6$~mm, acute angle of about $51^{\circ}$, and obtuse angle of about $124^{\circ}$ is shown in Figure~\ref{fig8}.  As it is apparent from this image, the spacing between the lines seen through the drop over the obtuse angular region {\em decreases\/} as they approach the vertex, while the spacing between the lines seen through the drop over the acute sector remains unchanged.  Note that only a few millimeters near the vertex should be taken into account while viewing this figure since the inverse mean curvature $R$ for water drops not distorted by gravity is of this order of magnitude.  At higher distances gravitational effects cannot be neglected while calculating the surface shape.  

This observation agrees favorably to the result of our calculation, as it can be seen from an argument based on geometrical optics.  Indeed, for the dimensions of the optical image along the bisector only radial curvature is important, and one can write the following approximate expression for the linear magnification by the drop in the $\hat r$ direction:
\begin{equation}
m = \frac 1 {1 + s (n-1) c_{rr}}
\end{equation}
where $n$ is the index of refraction of the liquid the droplets are made of\footnote{This expression assumes the horizontalness of the drop.}.  According to our result, for obtuse angles the curvature diverges and the magnification should go to zero as the vertex is approached; for acute angles both quantities remain finite.  Thus, qualitative validity of our result is confirmed by the simple demonstration described above.  Similar behavior should hold for the light reflected off the surface of the droplet because curvature is equally important for both such phenomena.

\begin{figure}
\begin{center}
\includegraphics{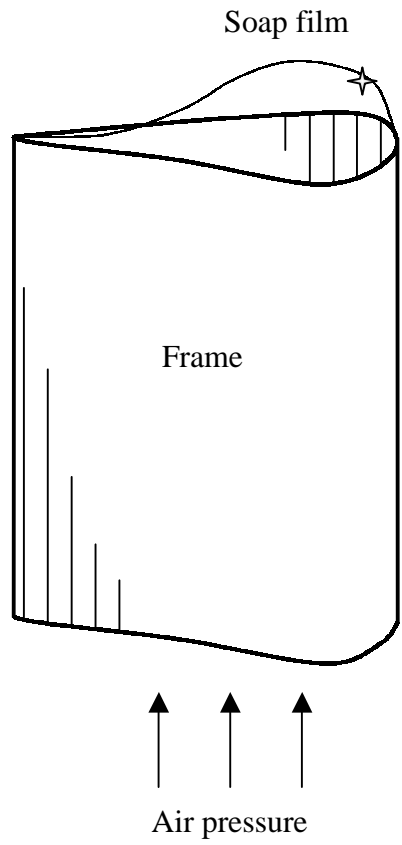}

\caption{Pressurized soap film realization.}
\label{fig9}
\end{center}
\end{figure}

Another possible system to test the predictions of our study is a pressurized soap film.  As shown in Figure~\ref{fig9}, a soap film on a wedge-shaped frame with an applied constant pressure difference across it will have constant mean curvature.  Thus, it is described by our formalism.  The only difference between such a film and a liquid drop is that the film has two surfaces, and therefore the applied $\Delta p$ must be two times as much as $\Delta p$ in equation~(\ref{laplace}).  Once such a surface is produced, it can be made as big as necessary for experimental convenience, since gravitational effects are virtually absent for this realization. 

As these examples illustrate, the change in behavior on going from acute to obtuse planar angular regions can show up in concrete ways.  We suspect that further differences will emerge as capillary flow and evaporation properties of these sector-shaped liquid interfaces are explored.

\vspace{3ex}

{\small The authors are grateful to Leo Kadanoff for an insightful discussion, to Joseph Keller for a useful mathematical clarification, to Sidney Redner for pointing out the analogous phenomena in hydrodynamics and diffusion, and to Itai Cohen for his help with photographing the drops.  This work was supported in part by the MRSEC Program of the National Science Foundation under Award Number DMR-9808595.}

\section*{Appendix}

Here we present some details on how we obtain expressions~(\ref{solution0}) and (\ref{solution1})-(\ref{solution4}).  One starts from the substitution of $ z(r,\phi)= r^\nu \Phi_\nu (\phi) $ (the first term of the expansion~(\ref{series})) into eq.~(\ref{exact}) and obtains the following equation for $\Phi_\nu (\phi)$:
$$ r^\nu\left(P[\Phi_\nu]+r^{2\nu-2}Q[\Phi_\nu]\right)+r^2\left(1+r^{2\nu-2}R[\Phi_\nu]\right)^{3/2}=0 \label{zeroth} $$
where $$P[\Phi_\nu]=\Phi_\nu^{\prime\prime}+\nu^2\Phi_\nu$$  $$Q[\Phi_\nu]=\nu^2\Phi_\nu^2\Phi_\nu^{\prime\prime}+\left(\nu-\nu^2\right) \Phi_\nu\left(\Phi'_\nu \right)^2+\nu^3\Phi_\nu^3$$  $$R[\Phi_\nu]=\left(\Phi'_\nu\right)^2+\nu^2\Phi_\nu^2$$
Considering all possible values of $\nu$, leaving only main terms in $r$ (the smallest powers of $r$) and solving for $\Phi_\nu(\phi)$ in each case, one arrives at the following set of solutions (only even terms are shown due to the symmetry of the problem): 
$$ \Phi_\nu(\phi)=\left\{\begin{array}{ll}
C\cos^\nu \phi &\qquad\qquad\mbox{if}\quad 0<\nu<1 \\ \\
C\cos\phi &\qquad\qquad\mbox{if}\quad \nu=1 \\ \\
C\cos\nu\phi &\qquad\qquad\mbox{if}\quad 1<\nu<2 \\ \\
C\cos 2\phi - \frac 14 &\qquad\qquad\mbox{if}\quad \nu=2 \\ \\
\mbox{no solution} &\qquad\qquad\mbox{if}\quad \nu>2
\end{array}\right. $$ 
($C$ is independent of $r$ and $\phi$ everywhere but arbitrary otherwise.)
Obviously, the first two options can not satisfy boundary conditions $\Phi_\nu(-\alpha/2)= \Phi_\nu(\alpha/2)=0$ for angles $\alpha<\pi$, and thus the ``not horizontal'' solutions with $\nu \le 1$ are naturally eliminated.  In cases 3 and 4 boundary conditions yield $\nu=\pi/\alpha$ and $C=1/(4\cos\alpha)$ respectively.  Thus, the main order result in the limit $r\ll 1$ is nothing but eq.~(\ref{solution0}):
$$ z(r,\phi)=\left\{\begin{array}{lll} \frac 14 r^2 \left(\frac{\cos 2\phi}{\cos\alpha}-1\right) &\qquad\qquad\mbox{if}\quad 0\le\alpha<\frac \pi 2 &\qquad(\nu=2)\\ \\ C r^{\pi/\alpha} \cos\frac{\pi\phi}{\alpha} &\qquad\qquad\mbox{if}\quad \frac \pi 2 <\alpha\le\pi &\qquad(\nu=\pi/\alpha) \end{array}\right. \label{solution5} $$

Then we proceed in exactly the same fashion to determine $\mu$ and to find $\Phi_\mu(\phi)$ by employing just calculated main order result.  Substitution of the first two terms of the expansion~(\ref{series}) into equation~(\ref{exact}) yields the following equation for $\Phi_\mu(\phi)$:
$$ r^\nu\left(P[\Phi_\nu]+r^{\mu-\nu}U[\Phi_\mu]+r^{2\nu-2}Q[\Phi_\nu]+r^{\nu+\mu-2}V[\Phi_\nu,\Phi_\mu]+O(r^{2\mu-2})\right)\,+ $$
$$ +\,r^2\left(1+r^{2\nu-2}R[\Phi_\nu]+r^{\nu+\mu-2}W[\Phi_\nu,\Phi_\mu] +O(r^{2\mu-2})\right)^{3/2}=0 \label{first} $$
where $\Phi_\nu(\phi)$ is already known, $P$, $Q$, $R$ are the same as in the equation for $\Phi_\nu(\phi)$, 
$$U[\Phi_\mu]= \Phi_\mu^{\prime\prime}+\mu^2\Phi_\mu$$ 
and expressions for $V$ and $W$ are irrelevant to any final results.  Similar to the previous case analysis, including thorough consideration of all possible cases for values of $\mu$, neglect of the terms of the order higher than the first correction in $r$ and application of symmetry arguments and boundary conditions to the solutions, leads to the first-order result~(\ref{solution1})-(\ref{solution4}).

Obviously, the procedure of building the next term of power series for the solution of eq.~(\ref{exact}) can be repeated up to an arbitrary order.

\end{document}